\documentclass[aps,pra,twocolumn,longbibliography,superscriptaddress,final,floatfix,
11pt,tightenlines]{revtex4-2}

\usepackage{amsfonts}
\usepackage{amsmath}
\usepackage{amssymb}
\usepackage[version=4]{mhchem}
\usepackage{subfigure}
\usepackage{latexsym}
\usepackage[export]{adjustbox}
\usepackage{array}
\newcolumntype{P}[1]{>{\centering\arraybackslash}p{#1}}
\newcolumntype{M}[1]{>{\centering\arraybackslash}m{#1}}
\usepackage{caption}
\usepackage[colorlinks,citecolor=red,urlcolor=blue,bookmarks=false,hypertexnames=true]{hyperref} 
\usepackage[usenames, dvipsnames]{color}
\usepackage[svgnames]{xcolor}
\usepackage{color}
\usepackage{bm}
\usepackage{relsize}
\usepackage{mathtools}
\usepackage{float}
\usepackage{subfloat}
\usepackage{dcolumn}
\usepackage{epsfig}
\usepackage{graphicx}
\usepackage{multirow}
\usepackage{makecell}
\usepackage{mathtools}
\usepackage[bbgreekl]{mathbbol}

\usepackage{footnote}
\usepackage{balance}

\usepackage{mathrsfs}
\usepackage{natbib}
\begin{document}

\title{Decoherence of a quantum magneto-oscillator : Effect of field-bath interaction}
 
\author{Koushik Mandal}
\affiliation{School of Applied Science and Humanities, Haldia Institute of Technology, Haldia-721657, India} 
\author{Suraka Bhattacharjee}
\thanks{Corresponding Author \\
Email: surakabhatta@gmail.com}
\affiliation{SASTRA Deemed University, Thirumalaisamudram, Thanjavur-613401, India}    
\author{Supurna Sinha}
\affiliation{Raman Research Institute, Bangalore-560080, India}
\date{\today}
\begin{abstract}
We investigate the quantum dissipative dynamics of a charged magneto-oscillator (CMO) coupled to a heat bath of harmonic oscillators in the presence of an additional field-bath interaction. The problem is formulated within the framework of the generalized quantum Langevin equation, where the external confining potential modifies the random force correlations and the memory kernel of the environment
and the spring constant. Using the corresponding non-Markovian master equation, we examine the influence of the field-bath coupling on the temporal decay of the reduced density matrix of the system. We show that the additional interaction leads to an enhancement of the decoherence rate by altering the dissipative response of the bath. Thus our results shed light on the role of field-bath coupling in the quantum to classical transition of a CMO. The resulting fluctuation-dissipation relation is analyzed, and the position autocorrelation, position-velocity correlation and velocity autocorrelation are obtained explicitly. The dependence of these experimentally accessible quantities on the field-bath coupling parameter is discussed. 
We outline an experimental proposal for testing our theoretical predictions. The study sheds light on reservoir 
engineering which is central to quantum technologies. 


\end{abstract}
\maketitle
\section{Introduction}
  The study of quantum Brownian motion (QBM) is central to modern quantum statistical physics, providing a fundamental framework for understanding how microscopic quantum systems interact with their environments \cite{Fisher1985,Eisert2002,Blume2008,Hu1993,Bauer2014,Pucci2013,Cobanera2016}. While classical Brownian motion describes the dynamics of particles coupled to a thermal bath, its quantum analogue captures much richer physics such as 
  decoherence of an open-quantum-system (OQS) and quantum technologies. Over the last century, QBM has evolved from a conceptual extension of classical stochastic theory into a mature discipline with rigorous mathematical foundations and broad applicability across condensed-matter physics, quantum optics, nanomechanics, and the quantum information domain.\\ 
  A charged particle trapped in a harmonic (or anharmonic) potential in an external magnetic field and coupled to a heat bath of oscillators undergoes quantum Brownian motion. Non-Markovian dynamics in open quantum systems has been developed via a systematic framework of dissipation theories and a variety of systems have been analyzed within this framework \cite{jchemone,jchemtwo,jchemthree}.
Further, the non-Markovian quantum dynamics of a charged magneto-oscillator (CMO)  coupled to a heat bath has been analysed for a variety of system-bath couplings like the CMO coupled to a heat bath via  position coupling \cite{suraka_pramana,decoherence1,asam_cmo_decoherence}, momentum coupling \cite{suraka_momentum,malay_2011} and via position-momentum coupling \cite{decoherence2}.\\
A charged magneto-oscillator undergoes decoherence (which results in a destruction of coherence responsible for quantum superposition) due to its coupling to the environment and the decoherence rate is affected by a number of factors like the cyclotron frequency ($\omega_c$), temperature or other parameters such as anharmonicity \cite{asam_cmo_decoherence, decoherence1,decoherence2,anharmonic_decoherence}. 
The decoherence rate goes up with the increase in temperature and is reduced as one increases the strength of the magnetic field. Non-Markovianity results in an information backflow which can be probed by studying the time evolution of the heating function \cite{decoherence1}. \\
 One of the questions of interest in recent years has been to understand 
the role of the coupling between the external potential and the bath of oscillators in influencing
the qualitative and quantitative behaviour of the memory kernel governing the dissipative behaviour 
of a Brownian particle. In \cite {pedro_2023,pedro_2024}, the author investigated a Brownian oscillator in the presence of a field-bath interaction alongside other standard interactions. Extending this analysis to a CMO in the same physical context offers a novel and compelling direction for exploration.
In a significant molecular dynamics simulation Daldrop et al. \cite{daldrop_2017} showed that for methane in a Lennard-Jones (LJ) water bath subjected to a harmonic potential, the kernel depends on both time and field strength because the friction is not a constant but is a function of the latter.
In recent works of  Vroylandt and Monmarche \cite{vroylandt_2022} they have applied the Zwanzig-Mori \cite{fukui1971,grabert1978,widder_2025} formalism to derive the generalised Langevin equation in terms of a nonlinear effective mean force and a friction linear in the velocity with a position dependent kernel. The MD simulation of two Lennard-Jones dimers \cite{straub1990} confirms the prediction. Furthermore,  Straub et al. \cite{straub1990} had shown using molecular dynamics, that the memory kernel is position dependent for a pair of particles in a simple liquid. Thus, for a parabolic potential, the simulation results indicate that the bath Hamiltonian must have a term involving an interaction of the external field with the reservoir degrees of freedom to obtain a more realistic physical description of the phenomenon. This motivates our present work in going beyond earlier works in investigating the role of the coupling between the external harmonic potential and the bath of oscillators in controlling the process of decoherence of a charged particle in a magnetic field.\\
In this paper we consider a charged magneto-oscillator (CMO) weakly interacting with a bath of harmonic oscillators via a position-position coupling. In addition, we introduce a field-bath interaction through a harmonic oscillator potential which is tunable via a parameter $\lambda$. We are interested in investigating the effect of this field-bath interaction on the rate of decoherence. 
We notice that the field-bath interaction enhances decoherence of the CMO. 
We detail our findings in subsequent sections. \\
The paper is organised as follows:  In section-2, we develop the quantum Langevin equation (QLE)
 for a CMO coupled to an environment of harmonic oscillators and in the presence of a field-bath interaction. We elaborate on the fluctuation-dissipation relation in section-3. Using the properties of the random force, the correlation functions (position-position, position-velocity and velocity-velocity) are derived in section-4.
The non-Markovian master equation is set up for the CMO and using this, the dynamics of the reduced density matrix (RDM) is derived in section-5. In section-6, the results of our investigation are summarised as plots studying the time variations of the correlation functions for several trapping frequencies and their variations with $\lambda$ are studied in an attempt to understand how the field-bath interaction 
affects the damping of these correlation functions. The evolution of the off diagonal elements of the reduced density matrix with time is plotted to study the effect of the field-bath interaction on decoherence. 
We present an experimental proposal for testing the predictions of our results in section-7. Finally, we conclude with some discussions and concluding remarks in section-8.
\section{Quantum Langevin Equation}
We consider a charged quantum Brownian particle of mass $M$ and charge $e$ coupled to a bath of $N$-harmonic oscillators at an arbitrary temperature $T$. This system is subjected to an external magnetic field $\vec{\mathcal{B}}$ along the $z$-axis. Thus the motion of the charged particle is confined to the two-dimensional $x-y$ plane. In addition, the Brownian particle is trapped in a harmonic potential.\\
The Hamiltonian for this system linearly coupled to a bath 
is given by \cite{pedro_2024}:
\begin{align}
   &\mathcal{H} = \frac{1}{2M}\left[\vec{p}-\frac{e}{c}\vec{A} \right]^{2} + V(r)+ \notag \\
   &\sum_j \left[ \frac{\vec{p}_j^2}{2m_j}+\frac{1}{2}m_j \omega_j^2 (\vec{q}_j-\vec{r})^2\right] + \sum_{j} \frac{\lambda}{2} M\omega_{0}^{2}q_{j}^{2} \label{Hamiltonianone}
\end{align}
The Hamiltonian in Eq.(\ref{Hamiltonianone}) can be explicitly written as:

\begin{equation}
    \mathcal{H} = \mathcal{H}_{S} + \mathcal{H}_{E} + \mathcal{H}_{SE} + \mathcal{H}_{P}
    \label{Hamiltonian}
\end{equation}
with
\begin{align}
    \mathcal{H}_{S} =& \frac{1}{2M}\left(\vec{p}-\frac{e\vec{\mathcal{A}}}{c}\right)^{2} + V(r) ,\label{syshamiltonian}\\
     \mathcal{H}_{E} =& \sum_{j} \frac{p_{j}^{2}}{2m_{j}} +\frac{1}{2}m_{j}\omega_{j}^{2}q_{j}^{2}, \label{envhamiltonian}\\
     \mathcal{H}_{P} =& \sum_{j} \frac{\lambda}{2} M\omega_{0}^{2}q_{j}^{2} \label{couphamiltonian}
\end{align}

where, $\mathcal{H}_{S}$, $\mathcal{H}_{E}$ and $\mathcal{H}_{P}$ are respectively the Hamiltonian of the system, the environment and the additional harmonic potential term associated with the bath particles.  $\vec{\mathcal{A}}$ $[(-\mathcal{B}y/2,\mathcal{B}x/2,0)]$ is the vector potential pertaining to the 
applied magnetic field $\vec{\mathcal{B}}$ and $\vec{p}$, $\vec{r}$ are respectively the momentum and the position coordinate of the Brownian  particle.
$\vec{p}_{j}$, $\vec{q}_{j}$, $m_{j}$ and $\omega_{j}$ are respectively the
momentum coordinate, the position coordinate, the mass and the frequency
of the $j$-th bath oscillator. \\
We consider a position-position coupling and the corresponding particle-bath interaction is  modeled as follows: 
\begin{equation}
    \mathcal{H}_{SE} = x \otimes \sum_{j} c_{j}q_{jx} + y \otimes \sum_{j} c_{j}q_{jy} \label{sys-envhamiltonian}
\end{equation}
where the coupled $x$ and $y$ coordinates are monitored by the environment. The system is linearly coupled to the environment via the position coordinate and $c_{j}$ is the coupling constant.  $q_{jx}$ and $q_{jy}$ are respectively the $x$ and $y$ components of $\vec{q}_{j}$.\\
Here, the coupling constant $c_j$ is given by $c_j=\frac{1}{2}m_j \omega_j^2$, corresponding to the Hamiltonian in Eq.(\ref{Hamiltonianone}).
The potential $V(r)$ is of the following form:
\begin{equation}
    V(r)= \frac{1}{2}M\omega_0^2\left[(x^{2}+y^{2}) \right]
\end{equation}
where $\omega_0$ is the frequency of the harmonic oscillator.
Notice, that in Eq.(1), following the standard procedure in open quantum system studies, we have introduced a counter term of the form $\frac{1}{2}m_j\omega_j^2r^2$ to compensate for the renormalisation of the system potential. \\
A tuning parameter $\lambda$ has been introduced to set the strength of the parabolic potential associated with the bath. This measures the extent to which the external harmonic potential influences the bath and thus 
captures the field-bath interaction which is the central focus of this paper.\\
We solve the system equations of motion in the Heisenberg picture ($\Dot{\mathcal{O}} = \frac{1}{i\hbar} [\mathcal{O},\mathcal{H}]$). The equations of motion for the CMO-coordinates are \cite{cohen2019}
\begin{align}
    \Dot{\Vec{r}}&= \frac{1}{M}\bigg(\Vec{p}-\frac{e\Vec{\mathcal{A}}}{c}\bigg)\\
    \Dot{\Vec{p}}&= -\Vec{\nabla}_r V(r) + \sum_{j}m_j \omega_j^2 (\Vec{q}_j -\Vec{r}) + \frac{e}{c} (\Vec{v}\times \mathcal{\Vec{B}}) \nonumber\\ 
    &+ \frac{i\hbar e}{2Mc}\big[\Vec{\nabla}\big(\ \Vec{\nabla}.\Vec{\mathcal{A}}\big) - \nabla^2 \Vec{\mathcal{A}} \big]
    \label{Heisenberg_eq_CMO}
\end{align}
Notice that the fourth term in Eq.(\ref{Heisenberg_eq_CMO}), reduces to zero
as the current density $\Vec{\mathcal{J}}= \frac{c}{4\pi} (\Vec{\nabla}\times \Vec{\mathcal{B}})= \frac{c}{4\pi} [\Vec{\nabla}(\Vec{\nabla}.\Vec{\mathcal{A}})-\nabla^{2}\Vec{\mathcal{A}}]$ produced by the external magnetic field pertains to a current source which is situated outside the domain of the moving charge. So the last term in Eq.(\ref{Heisenberg_eq_CMO}) vanishes and by rewriting the gradient of the harmonic potential as $M\omega_0^2 \vec{r}$ \cite{ford_1990} we get:
\begin{equation}
    M\Ddot{\Vec{r}} = -M\omega^2_0 \Vec{r} (t) + \frac{e}{c} (\Vec{v}\times \mathcal{\Vec{B}}) + \sum_{j}m_j \omega_j^2 (\Vec{q}_j -\Vec{r})  
    \label{r_CMO}
\end{equation}
The equations of motion of the j-th bath particle are
\begin{align}
    \Dot{\vec{q}}_j = \frac{\vec{p}_j}{m_j}
 \label{eqoma}
\end{align}
\begin{align}
    \Dot{\vec{p}}_j = -D_{j} \vec{q}_{j} + m_{j}\omega_{j}^{2} \vec{r}(t)
    \label{eqomb}
\end{align}
where $\Dot{\vec{p}}_j= m_{j} \Ddot{\vec{q}}_j$ and 
\begin{align}
D_{j}= (m_{j}\omega_{j}^{2}+\lambda M \omega_{o}^{2}) \label{Dj}
\end{align}
From Eqs.(\ref{eqoma}) and (\ref{eqomb}), we get the equation of motion for the bath coordinate as:
\begin{align}
    \ddot{\vec{q_j}}(t)+(D_j/m_j)\vec{q}_j(t)-\omega_j^2\vec{r}(t)=0
\end{align}

Now the solution of the bath coordinate is
\begin{align}
    \vec{q_{j}}(t) &=  \vec{q_{j}^h} (t) + \frac{\omega_{j}^{2}}{\tilde{\omega}_{j}^{2}} \vec{r}(t) \nonumber \\
    &- \frac{\omega_{j}^{2}}{\tilde{\omega}_{j}^2}\int_{-\infty}^{t} dt' \cos [\tilde{\omega}_{j}(t-t')] \Dot{\vec{r}}(t') 
    \label{bath_solution}
\end{align}
where, we define the modified bath frequency $\tilde{\omega}_j$ as \cite {pedro_2023}.
\begin{equation}
    \tilde{\omega}_j^2= \frac{D_j}{m_j}=\omega_j ^{2}+\lambda \frac{M}{m_j} \omega_o^{2}
    \label{omegajt}
\end{equation}
 Notice that the frequency of a bath particle gets modified for $\lambda \neq 0$. This leads to an increase in the effective spring constant with an increase in the field-bath interaction strength. \\
In Eq. (\ref{bath_solution}),
$\vec{q_{j}^{h}}(t)$ is the homogeneous solution of the bath coordinate given by:
\begin{align}
    \vec{q_j^h}(t)=\vec{q_j}(0)\cos( \tilde{\omega}_{j}t)+\frac{\vec{p_j}(0)}{m \tilde{\omega}_{j}}\sin(\tilde{\omega}_{j}t)
    \label{qhofj}
\end{align}
On eliminating the bath degrees of freedom from Eq.(\ref{r_CMO}) using Eq.(\ref{bath_solution}) we get:
\begin{align}
 &M\Ddot{\vec{r}}=-M \omega_0^2 \vec{r}+\sum_j m_j \omega_j^2 \bigg[\vec{q}_j^h(t)+\frac{\omega_j^2}{\tilde{\omega}_j^2} \vec{r}(t)- \notag \\
& \frac{\omega_j^2}{\tilde{\omega}_j^2}\int_{-\infty}^t dt' \cos[\tilde{\omega}_{j}(t-t')] \dot{\vec{r}}(t')-\vec{r}\bigg] +\frac{e}{c}(\vec{v} \times \vec{B}) \label{Langevinsteps}
\end{align}
Simplifying Eq.(\ref{Langevinsteps}), we get:

\begin{align}
    &M\Ddot{\vec{r}}+\sum_j\frac{m_j\omega_j^4}{\tilde{\omega}_j^2}\int_{-\infty}^t dt' \cos[\tilde\omega_j(t-t')] \dot{\vec{r}}(t')- \notag \\
    &\sum_j m_j \omega_j^2 \vec{q}_j^h (t)+
    \vec{r}(t)\bigg[M \omega_0^2-\sum_j  \frac{m_j \omega_j^4}{\tilde{\omega}_j^2}- \notag \\& m_j \omega_j^2 \bigg] -
    \frac{e}{c}(\vec{v} \times \vec{B})=0 \label{Langevinsteps1}
\end{align}
Finally, we arrive at the generalized quantum Langevin Equation (QLE) for the CMO as 
\begin{align}
    M\Ddot{\vec{r}}-\frac{e}{c}(\Vec{v}\times\Vec{\mathcal{B}})+ &\int_{-\infty}^{t} \mu(t-t') \Dot{\vec{r}}(t') dt'  \notag\\
    &+M \Omega_{o}^{2} \vec{r}(t) =\mathcal{\vec{F}}(t)
    \label{QLE}
\end{align}
where, $\mu$(t) is the memory kernel

\begin{equation}
    \mu(t-t') = \sum_{j} \frac{m_{j}\omega_{j}^{4}}{\tilde{\omega}_j^2}\cos [\tilde{\omega}_{j}(t-t')] 
    \label{memory_function}
\end{equation}
and $\mathcal{\vec{F}}(t)$ is the random force as discussed in detail in the next section.
\begin{equation}
    \mathcal{\vec{F}}(t) = \sum_{j} m_{j}\omega_{j}^{2}\vec{q_{j}^{h}}(t)
    \label{force_function}
\end{equation}
In Eq.(\ref{force_function}), $\mathcal{\vec{F}}(t)$ depends on $\tilde\omega_j$ via $\vec{q_{j}^{h}}(t)$ (See Eq.(\ref{qhofj})). Notice that the spring constant $\omega_0$ in Eq.(\ref{Langevinsteps1}) (the fourth term on the L.H.S.) gets renormalized in terms of the modified bath frequency $\tilde{\omega}_j$. We denote the effective spring constant by $\Omega_{o}$ which is given by \cite{pedro_2023}:

\begin{align}
    &\Omega_{o}^2 = \omega_0^2\Big(1+\lambda\sum_{j} \frac{\omega_{j}^{2}}{\tilde{\omega}^{2}_{j}}\Big)= \omega_0^2\Big(1+\lambda N_J\Big) \label{modifiedfreq}
    \end{align}
    where, $N_J=\sum_{j} \frac{\omega_{j}^{2}}{\tilde{\omega}^{2}_{j}}$.\\
Notice that $\Omega_{o}$ is the effective spring constant experienced by the CMO in the presence of the field-bath interaction. A non-zero $\lambda$ modifies the bare spring constant of the CMO. For $\lambda=0$ one recovers the original spring constant. 
We would like to point out some interesting features of the QLE for the CMO in the presence of the field-bath interaction, which are distinct from what one finds for the CMO in the absence of such an interaction \cite{suraka_pramana,ford_1990,decoherence1,decoherence2}. These are : (i) The modification of the spring constant of the CMO which captures the effect of the field-bath interaction. This field-bath interaction physically captures a situation in which the bath particles also get affected by the harmonic motion of the CMO in the presence of the external potential $V(r)$. As we mentioned earlier, in the limit of $\lambda \to 0$, 
one recovers the QLE presented in \cite{decoherence1,suraka_longtime}. However, in our subsequent analysis, for the sake of analytical tractability we will focus attention on the large $\lambda$ regime to compute the correlation functions, force correlation and decoherence rate. \\
(ii) The memory kernel $\mu(t)$ (see Eq.(\ref{memory_function})) and the force function $\mathcal{\vec{F}}(t)$ (see Eq.(\ref{force_function})) both depend on $\tilde\omega_j$,
leading to 
nontrivial fluctuation and dissipation effects, distinct from what one finds in the absence of the field-bath interaction \cite{suraka_longtime}.  \\
(iii) This dependence of the memory kernel and the force function on $\lambda$ leads to an enhanced damping, which in turn leads to an increase in the decoherence rate (See plots in the Results section).
\section{Fluctuation Dissipation Theorem}  
Here we derive the FDT for the CMO which establishes a direct connection between the equilibrium fluctuations of the noise and the dissipative properties of the bath. As the heat bath is at a fixed temperature $T$ and is coupled to the CMO, we use a statistical average of the bath operators \cite{shankar2012}. Thus we have
\begin{align}
    \langle q_{j\alpha}(0) \rangle &= 0\nonumber\\
    \langle p_{j\alpha}(0) \rangle &= 0\nonumber\\
    \langle q_{j\alpha}(0)q_{k\beta}(0) \rangle &= 
    \frac{\hbar}{2m_j\tilde{\omega}_j} \coth \bigg( \frac{\hbar\tilde{\omega}_j}{2k_{B}T}\bigg) \delta_{jk}\delta_{\alpha \beta}\nonumber\\
    \langle p_{j\alpha}(0)p_{k\beta}(0) \rangle &= 
    \frac{\hbar m_j\tilde{\omega}_j}{2} \coth \bigg( \frac{\hbar\tilde{\omega}_j}{2k_{B}T}\bigg)\delta_{jk}\delta_{\alpha 
    \beta}\nonumber\\
    \langle q_{j\alpha}(0)p_{k\beta}(0) \rangle&=-\langle p_{j\alpha}(0)q_{k\beta}(0) \rangle= \frac{1}{2}i \hbar\delta_{jk}\delta_{\alpha 
    \beta}
    \label{canonical_average}
\end{align}
Now using, Eqs.(\ref{force_function}), (\ref{bath_solution}), (\ref{omegajt}), (\ref{qhofj}), and (\ref{canonical_average}), we arrive at the noise correlation function for the effective random force, corresponding to the CMO model with additional field bath interaction: 
\begin{align}
    &\frac{1}{2} \langle\lbrace \mathcal{F}_\alpha(t), \mathcal{F}_\beta(t') \rbrace \rangle = \nonumber \\
    &\hbar \delta_{\alpha,\beta} \sum_{j} \frac{m_j \omega_{j}^4}{\tilde{\omega}_j} \coth{\bigg(\frac{\tilde{\omega}_j}{\Omega}\bigg)} \cos[\tilde{\omega}_{j} (t-t'))]
    \label{force_correlation}
\end{align}
where $\alpha, \beta = \lbrace x,y,z\rbrace$,  $\Omega=\frac{2 k_B T}{\hbar}$ and $\tilde{\omega}_j$ is the modified bath frequency as mentioned above. The random force $\mathcal{F}(t)$ is a zero-mean Gaussian function which obeys the force correlation given by Eq.(\ref{force_correlation}).
Notice that the force correlation is dependent on the coupling between the external potential and the bath via 
$\tilde{\omega}_j$ which is dependent on the coupling $\lambda$ (See the expression for $\tilde{\omega}_j$ in Eq.({\ref{omegajt}})). In the limit of $\lambda \rightarrow 0$ we recover the usual force correlation where one does not take into account the role 
of the coupling between the external potential and the bath \cite{suraka_longtime}. \\
Here, we consider the number of bath oscillators to be infinite and consider the $M>>m$ limit (more precisely, $\lambda M/m >>1$) for simplifying the coth term in Eq.(\ref{force_correlation}), to render the calculations tractable  (see Appendix A),
\begin{align}
 \nu(t)=&\hbar \gamma M \bigg[\frac{\cos(\Lambda t)+\Lambda t \sin(\Lambda t)-1}{t^2}\bigg] \times \notag \\&\coth\left(\frac{\sqrt{\lambda}\tilde{m}\omega_0}{\Omega}\right) 
\label{noise_correlation}
\end{align}
where, the masses of the bath oscillators are considered to be nearly equal ($m_j \approx m$) and $\tilde{m}=\sqrt{M/m}$ \cite{pedro_2023,decoherence2}. Moreover, the dissipation constant $\gamma$ is related to $\mu(\omega)$ via $ \mu(\omega)=M \gamma $ for $0<\omega\leq\Lambda$, where $\Lambda$ is the upper cut-off frequency, corresponding to the abrupt cut-off model for an Ohmic spectral function \cite{decoherence1}.

\section{Correlation functions}
The motion of the CMO is confined to the $x-y$ plane as the applied magnetic field is directed along the $z$-axis. 
Now using the properties of the random force, one can find the position auto-correlation function (for $x$ and $y$ coordinates), velocity auto-correlation and the position-velocity correlation along the lines outlined in ref. {\cite{suraka_pramana}}.\\
The Fourier-space solutions of position coordinates (see Appendix-A) are used to define the position autocorrelation function as \cite{huang2008,hanggi2005} 
\begin{align}
    C_x(\omega)= \frac{1}{2} \langle \lbrace x(\omega) , x^{*}(\omega) \rbrace\rangle \\
     C_y(\omega)= \frac{1}{2} \langle \lbrace y(\omega) , y^{*}(\omega) \rbrace\rangle
 \end{align}
Now using the noise correlation function derived in Eq.(\ref{noise_correlation}), one gets $C_x (\omega) = C_y(\omega)$. \\

\begin{widetext}

\begin{figure}[H]
\centering
\hspace{-1.4cm}\includegraphics[scale=1.]{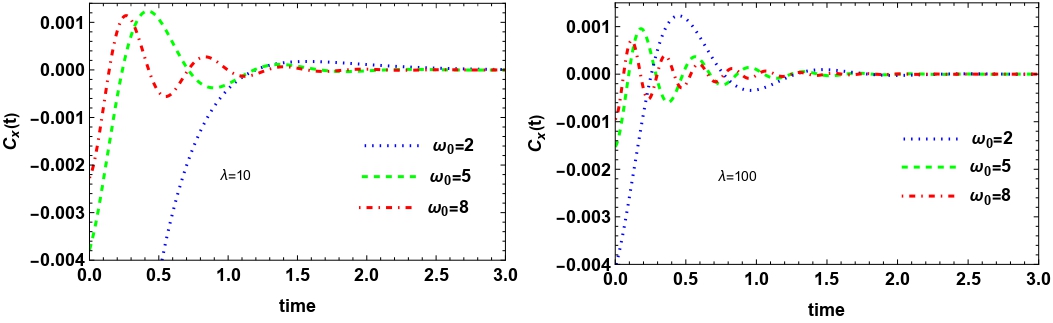}
\caption{ The position auto-correlation function versus time for different values of harmonic frequency $\omega_0$  and tuning parameter $\lambda$ with $\hbar=1$, $\omega_{c}=0.1$, $\gamma=5$, $\tilde{m}=100$, $\Omega=0.1$, $N_J=0.1$, $\Omega_0=\omega_0\sqrt{(1+\lambda N_J)}$.}
\label{posautocorr}
\end{figure}

\end{widetext}

\begin{widetext}

\begin{figure}[H]
\centering
\hspace{-2cm} \includegraphics [scale=1]{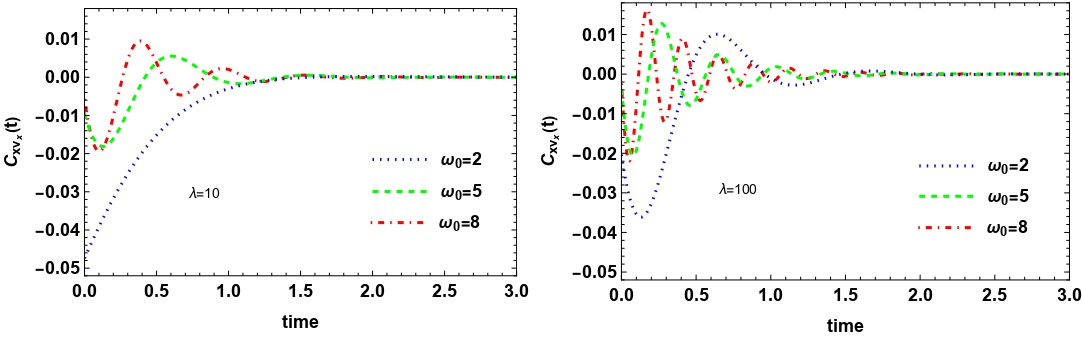}
\caption{ The position-velocity correlation function versus time for different values of harmonic frequency $\omega_0$ and tuning parameter $\lambda$ with $\hbar=1$, $\omega_{c}=0.1$, $\gamma=5$, $\tilde{m}=100$,  $\Omega=0.1$, $N_J=0.1$, $\Omega_0=\omega_0\sqrt{(1+\lambda N_J)}$.}
\label{posvelcorr}
\end{figure}

\end{widetext}

\begin{widetext}

\begin{figure}[H]
\centering
\hspace{-1.8cm}\includegraphics[scale=1.15]{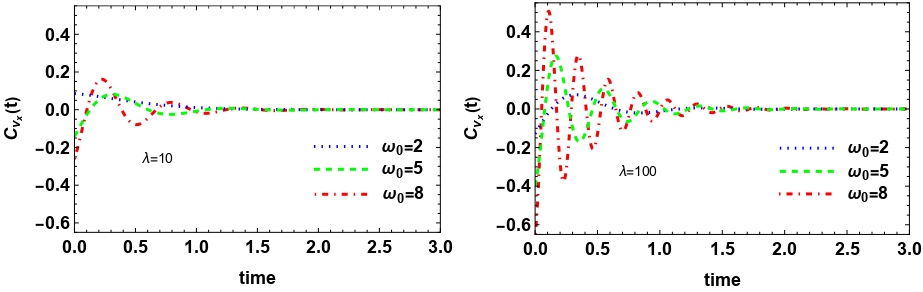}
\caption{The velocity auto-correlation function versus time for different values of harmonic frequency $\omega_0$  and tuning parameter $\lambda$ with $\hbar=1$, $\omega_{c}=0.1$, $\gamma=5$, $\tilde{m}=100$,  $\Omega=0.1$,  $N_J=0.1$, $\Omega_0=\omega_0\sqrt{(1+\lambda N_J)}$.}
\label{velautocorr}
\end{figure}

\end{widetext}

The position autocorrelation function is expressed in the time domain via the inverse Fourier transform as:
\begin{widetext}
\begin{align}
    C_x(t)= \frac{\hbar \gamma}{2 \pi M} \coth\left(\frac{\sqrt{\lambda}\tilde{m}\omega_0}{\Omega}\right)  
\int_{-\infty}^\infty d\omega \frac{\{(\omega-\frac{\Omega_0^2}{\omega})^2+\omega_c^2+\gamma^2\}e^{-i \omega t} }{\omega\{[(\omega-\frac{\Omega_0^2}{\omega})^2+\omega_c^2+\gamma^2]^2-4\omega_c^2(\omega-\frac{\Omega_0^2}{\omega})^2\}} 
\label{poscorr}
\end{align}
The position-velocity correlation function is defined as \cite{huang2008,suraka_momentum}
\begin{align}
C_{xv_{x}}(t)&=\frac{1}{2}\langle \left\lbrace x(t),v_x(0) \right\rbrace\rangle \nonumber\\
&=\frac{d}{dt}\frac{1}{2}\langle \left\lbrace x(t),x(0) \right\rbrace\rangle 
=\frac{d}{dt}C_{x}(t)
\end{align}
The velocity autocorrelation function is defined as \cite{huang2008,hanggi2005}
\begin{align}
C_{v_{x}}(t)&=\frac{1}{2}\langle \left\lbrace v_{x}(t),v_x(0) \right\rbrace\rangle \nonumber\\
&=-\frac{d^{2}}{dt^{2}}\frac{1}{2}\langle \left\lbrace x(t),x(0) \right\rbrace\rangle
=-\frac{d^{2}}{dt^{2}}C_{x}(t)
\end{align}
\end{widetext}
Notice that the correlation function $C_x(t)$ in Eq.(\ref{poscorr}) has an explicit dependence on $\lambda$ via the argument of the coth term and the modified harmonic frequency $\Omega_0$. We have calculated $C_x(t)$ numerically and plotted against time for different values of the parameter $\lambda$ (see Figs.(\ref{posautocorr})-(\ref{velautocorr})). 
\section{Non-Markovian Master Equation: Dynamics of the reduced density matrix}
We set up the Born-Markov master equation for a CMO linearly coupled to a heat-bath via position coordinates and in the presence of a field-bath interaction to study the dissipative and decohering dynamics of the system. Using the Born-Markov approximation, the Liouville-von Neumann equation for the system-density operator can be written as \cite{Schlosshauer2004,Schlosshauer2007}
\begin{align}
    \frac{\partial}{\partial t}\mathcal{\rho}_{s} (t) = -\frac{i}{\hbar} [ \mathcal{H}_{s},\rho_{s} (t) ] \nonumber\\
    -\frac{1}{\hbar^{2}} \lbrace   [S, B_{\alpha}\rho_{s} (t)] + [ \rho_{s} (t)C_{\alpha},S] \rbrace
    \label{system_update_equation}
\end{align}
we define the terms $B_{\alpha}$, $C_{\alpha}$  as follows
\begin{align}
    B_{\alpha}= \int_{0}^{\infty} d\tau \sum_{\beta} C_{\alpha\beta} (\tau) S_{\beta}^{(I)}(-\tau)\\
    C_{\alpha}= \int_{0}^{\infty} d\tau \sum_{\beta} C_{\beta\alpha} (-\tau) S_{\beta}^{(I)}(-\tau)
\end{align}
Here, the operators with the ${(I)}$ superscript pertain to the interaction picture, and the ones without a superscript pertain to the Schrodinger picture. We denote the system operator $S_{\alpha}(\tau)$ in the interaction picture as $S_{\beta}^{(I)}(\tau)$. $C_{\beta\alpha} (\tau)$ is a self-correlation function \cite{Schlosshauer2019quantum} for an operator $E$ evaluated over the initial state $\rho_{E}$ and is defined as $C_{\beta\alpha} (\tau)= \langle E_{\beta}(\tau) E_{\alpha}(0)\rangle$.\\
The non-Markovian master equation can also be derived for open systems where the system-environment coupling is weak \cite{Schlosshauer2007}. In the non-Markovian case, the form
of the master equation is the same, however, the operators $B_{\alpha}$ and $C_{\alpha}$ are time dependent.
\begin{align}
   B_{\alpha}= \int_{0}^{t} d\tau \sum_{\beta} C_{\alpha\beta} (\tau) S_{\beta}^{(I)}(-\tau)\\
   C_{\alpha}= \int_{0}^{t} d\tau \sum_{\beta} C_{\beta\alpha} (-\tau) S_{\beta}^{(I)}(-\tau)
\end{align}
It is to be noted that in the non-Markovian derivation of the master equation, the Born Approximation is assumed to hold true (weak coupling limit) \cite{Schlosshauer2004}.
$C_{\beta\alpha} (\tau)$ captures the extent of correlation between  
measurements of an observable $E$ at two different times
($\tau = t - t'$). The environmental correlation function provides a quantitative measure of the persistence of memory within the bath as it interacts with the system. The Markov approximation is invoked under the assumption that these correlation functions decay on a timescale much shorter than the intrinsic dynamical timescale of the system. We can find two different situations: (i) the self-correlation function $C_{\alpha\beta}(\tau )$ is peaked at $\tau = 0$ for
the Markovian scenario as a result of no memory retention, and (ii) the peak of $C_{\alpha\beta}(\tau)$ broadens
for the non-Markovian limit, implying the presence of finite memory retention.
However, the Born approximation is still considered to be valid, as the interaction between the system and the bath is weak such that the
system-environment density matrix is approximated as the tensor product of the system density matrix and the environmental density matrix ($\rho_{SE}(t) \approx \rho_{S}(t) \otimes \rho_{E}$): 
Now the non-Markovian dynamical equation for the CMO, within the Born approximation reduces to
\begin{widetext}
\begin{align}
    \frac{\partial}{\partial t}\mathcal{\rho}_s (t) = -\frac{i}{\hbar} [ \mathcal{H}_s,\rho_s (t) ]    -\frac{1}{\hbar} \int_{0}^{t} d\tau \lbrace \nu (\tau) [x, [x(-\tau),\rho_{s}(t)]] 
    -i\eta (\tau) [x, [x(-\tau),\rho_{s}(t)]] \nonumber\\ 
    - \nu (\tau) [y, [y(-\tau),\rho_{s}(t)]] 
    -i\eta (\tau) [y, [y(-\tau),\rho_{s}(t)]]\rbrace 
    \label{density_final}
\end{align}    
\end{widetext}
where the noise and dissipative kernels are respectively defined as:
\begin{align}
    \nu (\tau) &= \int_{0}^{\infty} d\tilde{\omega} J(\tilde{\omega}) \coth \bigg(\frac{\tilde{\omega}}{\Omega}\bigg) \cos(\tilde\omega\tau) \label{nutau} \notag\\
    \eta (\tau) &= \int_{0}^{\infty} d\tilde{\omega} J(\tilde{\omega})  \sin (\tilde\omega\tau)
\end{align}

The spectral density function for the  bath-oscillator $J(\omega)$ is defined as \cite{de2017,decoherence1}:

\begin{equation}
    J(\tilde{\omega}) = \sum_{j}\frac{c_{j}^2}{2m_{j}\omega_{j}}\delta(\tilde{\omega}-\omega_{j}) \label{spectral}
\end{equation}
For the Ohmic bath with abrupt cut-off model, the spectral density in Eq.(\ref{spectral}) is defined by:
\begin{align}
    J(\tilde{\omega})=\mu(\tilde{\omega}) \tilde{\omega}=M \gamma \tilde{\omega}
\end{align}
$\nu(\tau)$ in Eq.(\ref{nutau}) is solved in the limit of $M>>m_j$ and the result is displayed in Eq.(\ref{noise_correlation}).\\
We need to insert the solutions of particle coordinates ($x(\tau)$, $y(\tau)$) obtained by solving the system self equation of motion:
\begin{equation}
    M\Ddot{\Vec{r}}+M\Omega^2_0 \Vec{r} (t) -\frac{e}{c} (\Vec{v}\times \mathcal{\Vec{B}})  =0
    \label{r_system}
\end{equation}
Notice that the system self equation of motion in Eq.(\ref{r_system}) is similar to the system equations obtained in Ref.\cite{decoherence1, decoherence2}, however, the harmonic frequency is replaced by the modified frequency $\Omega_0$.\\
The solution of Eq.(\ref{r_system}) $x(\tau)$, $y(\tau)$ are obtained as follows:

\begin{widetext}
\begin{align}
    x(\tau)=&\frac{1}{2\sqrt{4 \Omega_0^2+\omega_c^2}}\bigg[\bigg \lbrace\left(-\omega_c+\sqrt{4 \Omega_0^2+\omega_c^2} \right) \cosh(A\tau)+ \notag \\
    &\left(\omega_c+\sqrt{4\Omega_0^2+\omega_c^2} \right) \cosh(B\tau) \bigg \rbrace X +\bigg \lbrace 2 \Omega_0^2  \left(\frac{\sinh(A \tau)}{A}-\frac{\sinh(B \tau)}{B} \right) \bigg \rbrace Y+\notag \\
   &\bigg \lbrace \left(\omega_c+ \sqrt{4 \Omega_0^2+\omega_c^2} \right)\frac{\sinh(A \tau)}{m A}+ \left(-\omega_c+\sqrt{4\Omega_0^2+\omega_c^2} \right)\frac {\sinh(B \tau)}{m B} \bigg \rbrace P_x +\notag \\ &\bigg \lbrace \frac{2}{m} \left( -\cosh(A \tau)+\cosh(B \tau)\right) \bigg \rbrace P_y\bigg] \label{couplx}\\
   y(\tau)=&\frac{1}{2\sqrt{4 \Omega_0^2+\omega_c^2}}\bigg[\bigg \lbrace 2 \Omega_0^2  \left(\frac{\sinh(B \tau)}{A}-\frac{\sinh(A \tau)}{B} \right) \bigg \rbrace X-\notag \\
   &\bigg \lbrace \left(-\omega_c+\sqrt{4 \Omega_0^2+\omega_c^2} \right) \cosh(A\tau)+\left(\omega_c+\sqrt{4\Omega_0^2+\omega_c^2} \right) \cosh(B\tau) \bigg \rbrace Y+\notag \\
   &\bigg \lbrace \frac{2}{m} \left(\cosh(A \tau)-\cosh(B \tau)\right) \bigg \rbrace P_x+\bigg \lbrace \left(\omega_c+ \sqrt{4 \Omega_0^2+\omega_c^2} \right)\frac{\sinh(A \tau)}{m A}+ \notag \\
   &\left(-\omega_c+\sqrt{4\Omega_0^2+\omega_c^2} \right)\frac {\sinh(B \tau)}{m B} \bigg \rbrace P_y\bigg]\label{couply}
   \end{align}
  \end{widetext}
where, 
   \begin{align}
    A=&\frac{1}{2}\Bigg[-2\Omega_{o}^{2}-\omega_c^{2}-\omega_c\sqrt{4\Omega_{o}^2+\omega_c^2}\Bigg]^{\frac{1}{2}}\\
    B=&\frac{1}{2}\Bigg[-2\Omega_{o}^{2}-\omega_c^{2}+\omega_c\sqrt{4\Omega_{o}^2+\omega_c^2}\Bigg]^{\frac{1}{2}}
  \end{align}
and $X = x(0)$, $Y = y(0)$, $P_x = M\Dot{x}(0)$, and $P_y = M\Dot{y}(0)$ are the initial position and momentum operators in the Schr\"{o}dinger picture. \\
Using $x(\tau)$, $y(\tau)$ as in Eqs.(\ref{couplx}),(\ref{couply}) and considering only the decoherence term, Eq.(\ref{density_final}) reduces to:
\begin{widetext}
\begin{align}
    \frac{\partial}{\partial t}\mathcal{\rho}_s (t) = -\frac{1}{\hbar} \int_{0}^{t} d\tau   \nu (\tau) F_{1}(\tau)[X, [X,\rho_{s}(t)]]  
-   \frac{1}{\hbar} \int_{0}^{t} d\tau   \nu (\tau) F_{1}(\tau)[Y, [Y,\rho_{s}(t)]]  \nonumber\\
-\frac{1}{\hbar} \int_{0}^{t} d\tau   \nu (\tau) F_{2}(\tau)[X, [Y,\rho_{s}(t)]]  
+   \frac{1}{\hbar} \int_{0}^{t} d\tau   \nu (\tau) F_{2}(\tau)[Y, [X,\rho_{s}(t)]]
\label{rhofinal}
\end{align} 
\end{widetext}
Here, the functions $F_1(\tau)$ and $F_2(\tau)$ depend on the modified harmonic frequency $\Omega_0$ and the cyclotron frequency $\omega_c$ as shown in Eqs.(\ref{F1}), (\ref{F2}):
\begin{widetext}
\begin{align}
    F_{1}(\tau)=& \frac{(-\omega_{c}+\sqrt{4\Omega_{o}^2+\omega_c^2})\cosh(A\tau)+(\omega_{c}+\sqrt{4\Omega_{o}^2+\omega_c^2})\cosh(B\tau)}{2\sqrt{4\Omega_{o}^2+\omega_c^2}} \label{F1}\\
    F_{2}(\tau)=&\frac{2\sqrt{2}(-B\sinh(A\tau)+A\sinh(B\tau))}{AB\sqrt{4\Omega_{o}^2+\omega_c^2}} \label{F2}
\end{align}
\end{widetext}

It is pertinent to mention here that the commutation brackets in the last two terms of Eq.(\ref{rhofinal}) represent the cross terms in the $x$ and $y$ directions which are equal to each other. Thus the last two terms in Eq.(\ref{rhofinal}) get cancelled \cite{decoherence2,anharmonic_decoherence} and the decoherence is controlled by the $x$ and $y$ spatial decohering terms.\\ 
Therefore, the off-diagonal elements of the system density matrix decay as
\begin{align}
    \rho_{s}  (x,y,x^{'},y^{'},t) = \rho_{s}  (x,y,x^{'},y^{'},0) \nonumber\\
    \exp{\big[- \int_{0}^{t} \mathcal{D}(t^{'}) dt^{'}\big]}
\end{align}  
where, 
\begin{equation}
    \mathcal{D}(t) = \Gamma(t) \big[ (\Delta x)^{2} + (\Delta y)^{2}\big] 
\end{equation}
with $\Delta x = x-x^{'}$ and $\Delta y=y-y^{'}$ and 
\begin{align}
    \Gamma (t) &=\frac{1}{\hbar}\int_{0}^{t} d\tau \nu(\tau) F_{1}(\tau) 
\end{align}
Here, we use the noise kernel in Eq.(\ref{noise_correlation}) to obtain the explicit form of $\Gamma (t)$. 
In the above derivation of the non-Markovian master equation we have 
confined to the weak coupling limit. There have been other treatments of non-Markovian 
systems \cite{jchemone,jchemtwo,jchemthree}  where the authors have developed a rigorous approach towards analysing such systems. \\

\section{Results}
In this section we analyze the behaviour of various experimentally observable quantities like the time evolution of the position and velocity autocorrelation functions, the position-velocity correlation functions and the off-diagonal elements of the reduced density matrix. 
\begin{widetext} 

\begin{figure}[H]
\centering
\hspace{-1.2cm}\includegraphics[scale=1]{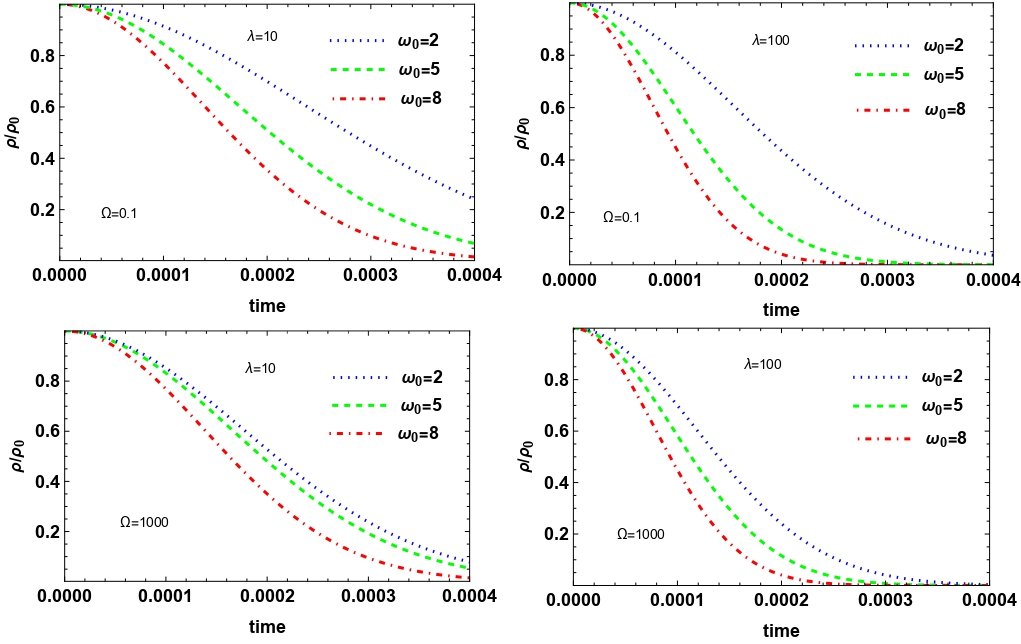}
\caption{The reduced density matrix ($\rho/\rho_0$) versus time for different values of the harmonic frequency $\omega_0$, tuning parameter $\lambda$ and temperature scale $\Omega$ with $\hbar=1$, $\omega_{c}=0.1$, $\gamma=1$, $\tilde{m}=100$, $N_J=0.1$, $\Omega_0=\omega_0\sqrt{(1+\lambda N_J)}$.}

\end{figure}

\end{widetext}
In Figs.($1$-$3$), one notices an oscillatory behaviour of the position autocorrelation, position-velocity and velocity autocorrelation functions. This oscillatory behaviour is associated with the cyclotron frequency $\omega_c$. However, the oscillations get damped away with time. The damping of the correlation functions gets enhanced with the increase in the strength of 
the modified harmonic frequency $\Omega_0$, which, in turn, is controlled by tuning the trap frequency $\omega_0$, according to Eq.(\ref{modifiedfreq}). 
Furthermore, we notice an enhanced rate of damping of these correlation functions with the increase in the coupling strength ($\lambda$) between the external harmonic trapping potential and 
the bath which also follows from the dependence of $\Omega_0$ on $\lambda$ via Eq.(\ref{modifiedfreq}) . This increased rate of damping with an increase in $\lambda$ and $\omega_0$ is physically expected due to the coupling of the external potential to the heat bath, contributing to an enhanced dissipative damping. \\
In Fig. 4, we have plotted the time variation of the off-diagonal elements of the reduced density matrix for different values of $\lambda$ and frequency $\omega_0$. The decay of the off-diagonal elements of the reduced density matrix indicates the loss of coherence and destruction of superposition of states leading to a quantum-to-classical transition in an open quantum system \cite{Schlosshauer2004, decoherence1}. Thus the plots exhibiting a faster decay of $\rho/\rho_0$ indicate an enhancement of decoherence with the increase in $\lambda$ and $\omega_0$, consistent with the faster rate of decay of the correlation functions with the increase in the strength of the harmonic frequency $\omega_0$ and its coupling to the heat bath (see Figs.($1$-$3$)).  Moreover, from Fig. 4, one notices a faster decay of coherence at higher temperatures, which signifies an earlier onset of quantum-to-classical transition in the higher temperature regime, as seen in our earlier works \cite{decoherence1,decoherence2,anharmonic_decoherence}. Thus a higher value of the coupling strength $\lambda$ leads to a stronger environmental coupling and a greater loss of information from the system, contributing to a faster decoherence with the increase in $\lambda$. A similar effect is noticed when the external harmonic frequency $\omega_0$ is large, leading to a stronger field-bath interaction (see Eq.(\ref{couphamiltonian})). 
In the next section we have discussed a possible 
experimental setup where our predictions can be tested. We expect the 
qualitative predictions made here, for instance, the trends with the variation of $\lambda$ and temperature, to be observable in experimental measurements. However, a detailed quantitative prediction lies beyond 
the scope of the present study.



\section{Experimental Proposal}
An experimental platform can be engineered to test the predictions of our theoretical model. Hybrid ion–atom trapping techniques provide a suitable route for confining the CMO. In particular, we consider a Penning trap, where stable confinement is achieved via the superposition of static quadrupolar electric fields (generated by ring and end-cap electrodes) and a uniform magnetic field \cite{Blaum2020}.\\
Dissipative cooling of the trapped ions is implemented using counter-propagating circularly polarised laser beams, analogous to a magneto-optical trap configuration \cite{ian_2008,camerer_2011}. This arrangement produces an effective optical molasses, which acts as a controllable Ohmic reservoir in correspondence with the bath assumed in our theoretical formulation \cite{optcont2022}. Under these conditions, a charged ion cloud confined in a harmonic potential and subjected to a magnetic field experiences decoherence due to its coupling with the engineered dissipative environment. The goal of this experimental setup is 
to check if the predictions stemming from our theoretical analysis are borne out. For instance, an increase in the decoherence rate with the increase in the field-bath coupling characterised by $\lambda$ can be tested via such a setup. The focus is 
more on the qualitative trends rather than quantitative comparison.
\begin{widetext}

\begin{figure}[H]
    \centering
    \includegraphics[width=\linewidth]{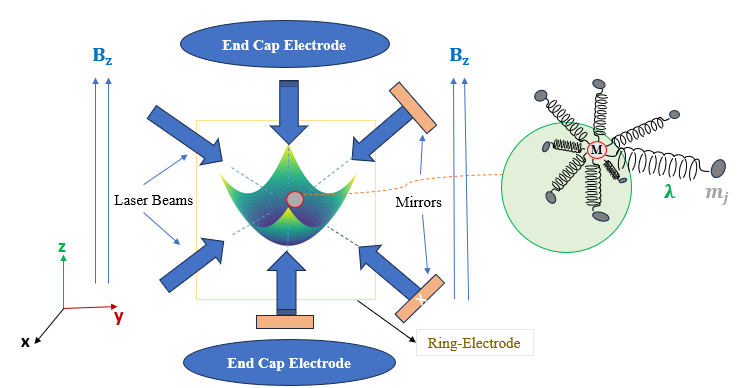}
    \caption{\textbf{Proposed experimental set up:} Our proposed experimental setup for a CMO in a Penning trap subjected to an additional field-bath interaction. The counter-propagating laser beams are used to cool the trapped ions and to provide optical molasses. An externally applied magnetic field is applied in the z-direction, as shown in the figure. The field-bath interaction is shown in the inset.}
    \label{expt_setup}
\end{figure}        
       
\end{widetext}
The proposed setup enables the control of the motion of the CMO via laser pulses, allowing direct access to quantum decoherence dynamics, via different experimental techniques, for example, Raman Ramsey fringe measurement in cold atom interferometry \cite{Hu,Black}. The field-bath interaction between the harmonic trap and the optical molasses can be implemented via a suitable feedback mechanism \cite{kim_2024}. 
The parameter $\lambda$ can be tuned by using the technique of species-specific optical lattices \cite{leblanc} which enables one to engineer specific lattice environments for the system and the bath atoms. Thus, while both the system and the bath particles experience harmonic confinement, the relative strength of the confinement can be tuned using this technique, enabling one to 
independently implement the confining potentials for the system and the bath atoms.
For consistency with our theoretical assumptions, the experimental parameters should satisfy: trap frequency $\omega_0/2\pi$ in MHz order, cyclotron frequency $\omega_c/2\pi$ in kHz order and the field-bath interaction tuning parameter $\lambda \sim 10$, ensuring the validity of the approximations used in our theoretical formulation. The bath temperature should be tunable across both regimes considered in our analysis, namely 
T $\sim$ mK (high-temperature limit) and 
T$\sim$ $\mu$K (low-temperature limit) respectively.\\
Decoherence times can be measured using Ramsey interferometry, with resolution optimised via appropriate control of the Rabi frequency and detuning. 
A setup similar to the one discussed in \cite{optcont2022} can be used to measure the correlation functions computed in this paper.  
A large mass ratio $\tilde m$ of the system particle mass to the bath particle mass can be 
realised by choosing a system consisting of a $Cs$ atom in a bath of $Li$ atoms for which the decoherence time is of the order of milliseconds to tens of milliseconds, under typical ultracold conditions \cite{tung_2013}.
\section{Discussion}
In this work, we have investigated the dissipative quantum dynamics of a charged magneto-oscillator interacting with a thermal reservoir in the presence of an additional coupling between the external confining field and the bath degrees of freedom. 
Our work follows on similar research works in the context of Brownian dynamics \cite{pedro_2023,pedro_2024} where 
the authors confine to a classical situation. We go beyond these studies in addressing quantum 
dissipative dynamics and decoherence. By extending the conventional system–reservoir model to include this field–bath interaction, we derived a modified generalized quantum Langevin equation in which both the effective restoring force and the memory kernel of the environment are renormalized by the coupling parameter. This demonstrates that the external trapping potential can influence the open-system dynamics not only through the system Hamiltonian but also indirectly through the statistical properties of the surrounding bath.

Using the reduced density matrix formalism, we analyzed the resulting decoherence dynamics of the oscillator and showed that the additional field–bath interaction strengthens the decay of quantum coherence. The enhancement of the decoherence rate originates from the modification of the bath-induced fluctuations and from the corresponding change in the dissipative response of the reservoir. In particular, the presence of the additional coupling effectively alters the temporal correlations of the environmental noise, resulting in a faster loss of coherence with the increase in the strength
of the coupling.

The correlation functions obtained in the present analysis further illustrate the influence of the modified reservoir structure on the observable dynamics of the system. The position autocorrelation, position–velocity correlation, and velocity autocorrelation functions all exhibit signatures of increased rate of damping as the field–bath interaction is increased. 
From the perspective of nonequilibrium statistical mechanics, the present results show that a relatively simple modification of the reservoir Hamiltonian can generate measurable changes in fluctuation and transport properties at the quantum level.

In future we would like to extend our analysis beyond an Ohmic bath model and explore other bath models, for instance Drude model and Lorentzian bath model to test the robustness of our results. We would also 
like to go beyond the realms of the approximations used in our analysis and explore a wider parameter range for 
the field-bath interaction and the mass ratio of the system and the bath particles. 

The predicted effects appear in experimentally accessible quantities and therefore the model can be tested in trapped-ion or hybrid quantum platforms where both magnetic confinement and 
dissipation can be realized under controlled conditions. Beyond the specific system considered here, the analysis suggests that coupling external control fields directly to environmental degrees of freedom may provide a general mechanism for manipulating irreversible quantum dynamics and therefore is expected to play an important role in the rapidly growing realm of quantum 
technology. 

In a broader context, our results indicate that environmental structure itself can serve as an active resource for controlling quantum coherence. This may be relevant for future studies of nonequilibrium open systems in which the interplay between confinement, dissipation, and memory plays an essential role. The present work therefore points toward a route for tailoring decoherence in mesoscopic quantum systems through controlled modification of the reservoir.
\section{Acknowledgement}
We thank Sanjukta Roy for valuable discussions on the experimental setup and feasibility of our predictions. We gratefully acknowledge the anonymous reviewers for their rigorous evaluation and insightful feedback, which substantially strengthened the analytical depth and clarity of this manuscript.
\section*{Appendix A}
The noise correlation function, corresponding to the random force is derived in Eq.(\ref{force_correlation}) as:
\begin{align}
    &\frac{1}{2} \langle\lbrace \mathcal{F}_\alpha(t), \mathcal{F}_\beta(t') \rbrace \rangle = \nonumber \\
    &\hbar \delta_{\alpha,\beta} \sum_{j} \frac{m_j \omega_{j}^4}{\tilde{\omega}_j} \coth{\bigg(\frac{  \tilde{\omega}_j}{\Omega}\bigg)} \cos[\tilde\omega_j (t-t')]
\end{align}
By incorporating the memory kernel, the above equation can be expressed as:
\begin{align}
    &\frac{1}{2} \langle\lbrace \mathcal{F}_\alpha(t), \mathcal{F}_\beta(t') \rbrace \rangle = \nonumber \\
    &\hbar \delta_{\alpha,\beta} \sum_j \mu(\omega_j) \tilde{\omega}_j \coth{\bigg(\frac{ \tilde{\omega}_j}{\Omega}\bigg)} \cos[\tilde\omega_j (t-t')]
\end{align}

Using Eq.(\ref{omegajt}) and replacing the summation by an integration in the limit of a large number of bath oscillators we get 

\begin{align}
    &\frac{1}{2} \langle\lbrace \mathcal{F}_\alpha(t), \mathcal{F}_\beta(t') \rbrace \rangle = \nonumber \\
    &\hbar \delta_{\alpha,\beta} \int_0^\infty \mu(\tilde\omega) 
 \tilde\omega\coth{\bigg(\frac{  \tilde{\omega}}{\Omega}\bigg)} \nonumber \\ 
     &\times\cos[\tilde \omega (t-t')] d\tilde\omega
    \label{noisecorrint}
   \end{align} 
In Eq.(\ref{noisecorrint}), we have considered the masses of the bath particles to be nearly equal ($m_j \approx m $) \cite{decoherence2} and $\tilde{m}=\sqrt{M/m}$.\\
 We consider the mass of the Brownian particle to be much larger than the  mass of the bath oscillators, i.e, $M >> m$, or more precisely, we consider  $ \lambda M/m >> 1$. In this limit the noise correlation function in Eq.(\ref{noisecorrint}) takes the form:

\begin{align}
  &\nu(t-t')= \hbar \gamma M \int_0^\Lambda \tilde \omega  \coth\left[\frac{\left(\sqrt{\lambda}\tilde{m}  \omega_0\right)}{\Omega}\right] \notag \\
  \times &\cos\left[\tilde\omega (t-t')\right] d\omega \label{forcecorrelationmod}
\end{align}
where, the memory kernel for the Ohmic bath takes the form $\mu(\omega)=M \gamma$ for $0<\omega\leq\Lambda$.\\
Solving the integration in Eq.(\ref{forcecorrelationmod}), in the large $\tilde{m}$ limit, one gets,
\begin{align}
 \nu(t)=& \hbar \gamma M  \bigg[ \frac{\cos(\Lambda t)+\Lambda t \sin(\Lambda t) -1}{t^2}\bigg] \times \notag \\
&\coth\left(\frac{\sqrt{\lambda}\tilde{m}\omega_0}{\Omega}\right) 
\label{noise_corr_final}
\end{align}
\section*{Appendix B}
 We have set up the QLE for the CMO (see Eq. (\ref{QLE}) in the main text). From the QLE, one can get the two coupled differential equations in $x$ and $y$ as 
 \begin{align}
     \Ddot{x}(t)-\omega_c \Dot{y} +&\frac{1}{M}\int_{-\infty}^{t} \mu (t-t^{'})\Dot{x}(t^{'})dt^{'} \notag \\
     & +\Omega_{o}^{2} x = \tilde{\mathcal{F}}_{x}(t)
     \label{forces}
     \end{align}
     \begin{align}
     \Ddot{y}(t)+\omega_c \Dot{x} +&\frac{1}{M}\int_{-\infty}^{t} \mu (t-t^{'})\Dot{y}(t^{'})dt^{'} \notag \\
     &+\Omega_{o}^{2} y = \tilde{\mathcal{F}}_{y}(t)
     \label{forcey}
 \end{align}
We define the cyclotron frequency as $\omega_c= \frac{e{B}}{Mc}$ and scale the force components by the particle mass $M$ 
as $\tilde{\mathcal{F}}_{x(y)}=\big({\mathcal{F}}/M\big)_{x(y)}$.
Now taking the Fourier transform on both sides of the above two coupled equations, we arrive at
\begin{align}
    [-\omega^2-i\omega \mathcal{K}(\omega)+\Omega_{o}^2] x(\omega)-i\omega_c\omega y(\omega)= \tilde{\mathcal{F}}_x(\omega)\label{c1}\\
    i\omega_c\omega x(\omega) + [-\omega^2-i\omega \mathcal{K}(\omega)+\Omega_{o}^2] y(\omega) = \tilde{\mathcal{F}}_y(\omega)
    \label{c2}
\end{align}
where, $\mathcal{K}(\omega)= \frac{\mu(\omega)}{M}$ and $\mu(\omega)= \int_{-\infty}^{+\infty} dt \mu (t) e^{i\omega t}$.  By solving Eq.(\ref{c1}) and (\ref{c2}), we get the Fourier space solution as: 
\begin{widetext}
\begin{align}
    x(\omega)&=\frac{(-\omega^2 +i\omega \mathcal{K}(\omega)+\Omega_{o}^2)\tilde{\mathcal{F}}_x(\omega)+ i\omega \omega_c \tilde{\mathcal{F}}_y(\omega)}{(-\omega^2+i\omega\mathcal{K}(\omega)+\Omega_{o}^2)^2 + (i\omega\omega_c)^2} \\
    y(\omega)&=\frac{-i\omega \omega_c \tilde{\mathcal{F}}_x(\omega)+ (-\omega^2 -i\omega \mathcal{K}(\omega)+\Omega_{o}^2)\tilde{\mathcal{F}}_y(\omega)}{(-\omega^2+i\omega\mathcal{K}(\omega)+\Omega_{o}^2)^2 + (i\omega\omega_c)^2}   
\end{align}
\end{widetext}
Making use of these solutions in $\omega$-space, we calculate the position autocorrelation, position-velocity and velocity autocorrelation functions (see main text). 

 
%
 \end{document}